\documentclass [12pt,a4paper]{article}
\usepackage[english]{babel}
\begin{document}
%\markright{Resonat4.tex\hfill\today}

\renewcommand{\thesection}{}

\renewcommand{\baselinestretch}{1.5}
\setcounter{secnumdepth}{0}

\begin{titlepage}

\begin{center}
{ \LARGE \textbf {Discovery of a Planar Waveguide for an X-Ray Radiation}}

\bigskip

V.K.~Egorov\footnote{E-mail: egorov@ipmt-hpm.ac.ru},
$^*$E.V.~Egorov

IPMT RAS, Chernogolovka, Moscow Dist, Russia

$^*$MEPhI, Moscow, Russia
\end{center}

\renewcommand{\abstractname}{}

\begin{abstract}

A simple model of X-Ray standing waves (XSW) formation in the slit
of a planar waveguide of X-Ray radiation beam  for the angle area
restricted by the critical total reflection angle is developed. It is shown that the model is true for a case of the Bragg reflection.
The conditions required for XSW to appear in the space between two
polished parallel plane plates are formulated and a slit size
interval conforming to these conditions is evaluated. A mechanism
of a XSW intensity decrease in a planar waveguide is proposed.
This mechanism explains a high efficiency of slitless collimator
application for the transportation of narrow X-Ray beams.

Some recommendations on the application of the planar X-Ray waveguide in
X-Ray structural and spectral studies of surface are presented.

\end{abstract}


\end{titlepage}

\section{Introduction}

A slit former of X-Ray beams with extensive plane restraints has long been
used as one of the basic components in X-Ray optics. A well-known example is
the Soller slit that contains a set of closely spaced thin parallel metallic
plates ensuring an X-Ray beam with a required divergence \cite{X-RAY1,PHYS1E}. Another example of
the unit is an aligned diffractometric slit 50~mm long with a clearance size
between the parallel quartz plates of 0.1~mm \cite{X-RAY1,RD63}.

These slits are sometimes used to form an X-Ray excitation beam in TXRF
spectrometry \cite{NIM1,ZAV1E}. For example, an X-Ray collimator formed by two quartz
parallel plates with a 0.1~mm clearance served as a double X-Ray beam
reflector to form an excitation beam for total X-Ray reflection fluorescence
analysis \cite{NIM1}.

In X-Ray diffraction practices the plane slit monochromators used for
the X-Ray line narrowing in an excitated beam has a wide application
\cite{APL1E,FIZTT1E}.
This devices designates as Bonse-Hart monochromators are distinguished
by multiple reflections within a groove cut into a large
single crystal or within a slit formed by two oriented monocrystal plates.

Lately, in addition to X-Ray planar extensive collimators characterized by
the visible size of a clearance between solid reflectors formed it, so
called slitless X-Ray collimators has come into use for X-Ray fluorescence
analysis \cite{IET2E,ELPROM1E,IZV1E,ADV1,POV1E}. The slitless collimator applied in those works was formed
by two quartz polished plates mated together. The clearance in it was formed
due to roughness and microsphericity of the plates, which size been made as
evaluation rather laborious and ambiguous. Experiments showed, that these
slitless collimators could transport X-Ray radiation over a distance of 100
mm without visible intensity decrease. Because the hypothesis multiple total
reflection of an X-Ray beam in the microclearance of a slitless collimator
failed to explain the experimental data obtained in \cite{IET2E,IZV1E}
, a hypothesis
of an X-Ray standing wave (XSW) formation in the microclearance was proposed
in \cite{ADV1,POV1E}. The present work is devoted to further development of this
hypothesis.

\section{Formation of an X-Ray standing wave upon specular reflection of a plane wave}

Assume that an electromagnetic monochromatic plane travelling wave
with the $\sigma $-polarization (i.e. $\vec E_0$ perpendicular to
the x-z plane in Fig.~1), wavelength $\lambda_0$ and wave vector
$k_0 = \frac{1}{\lambda_0}$ impinges on the boundary separating
two materials. If the materials have different refraction indices,
part of the wave energy is reflected and the remainder passes to
the second material or is refracted. An interference field appears
in the first material irrespective of the remainder value. The
interference field area depends on the width of an incident plane
wave and the incidence angle $\theta $. The intensity of the
interference area is directly determinated by the reflection
factor on the boundary between materials and peaks at the total
reflection of an incident radiation beam. Referring to Fig.~1a,
the incident and reflected travelling E-field plane waves can be
described as \cite{PRINCIPLES}:

\begin{equation}
\label{eq1}
\vec A_0 ( \vec r;t) = \vec E_0
e^{i\left[ {\omega t - 2\pi ( \vec k_x x - \vec k_z z )} \right]}
\end{equation}

\noindent
and

\begin{equation}
\label{eq2}
\vec A_R ( \vec {r};t) = \vec E_R
e^{i\left[ {\omega t - 2\pi ( \vec k_x x + \vec k_z z )} \right]}
\end{equation}

For the sake of convenience, let $z=0$ corresponds to the reflector surface.

By locating the intersections of the crests and troughs of the two
travelling plane waves in Fig.~1a, one can easily show that the
interference between the two coherent waves generates a standing
wave with planes of maximum and minimum intensity parallel to the
boundary surface. The period of the standing wave is defined by
expression \cite{PHYS2}:

\begin{equation}
\label{eq3}
D = \frac{\lambda}{2 \cdot \sin\theta}
\end{equation}

In a general case the amplitude relation between the incident and reflected
waves is described by the Fresnel equations \cite{PRINCIPLES}:

\begin{equation}
\label{eq4}
\left| {\frac{\vec {E}_{R} }{ \vec {E}_{0} }} \right|_{ \bot}  =
\frac{{\sin\theta - n \sin\varphi} }{{\sin\theta + n \sin\varphi} }
\end{equation}

\noindent where $\varphi $ is the refraction angle and $n$ is the
relative refraction index. The phase change of the electric vector
between the incident and reflected waves $\psi $ is defined by
expression:

\begin{equation}
\label{eq5}
\tan\frac{\psi}{2} = \frac{\sqrt {\cos^{2}\theta - n^2} }{\sin\theta}
\end{equation}

Equations (\ref{eq3}), (\ref{eq4}) and (\ref{eq5}) describe the reflection phenomenon for any
electromagnetic plane wave on a plane interface between two materials. They
can also be used to describe, as a first approximation, a total external X-Ray reflection on a vacuum -
material interface, which is usually represented as the specular reflection
of an X-Ray plane wave, for the sake of simplification%
\footnote{
This simplified representation disregards the Goos-Hanchen wave front
displacement at total wave reflection \cite{PLANAR}, and the radiation penetration to top layers of the material plate.}.
The interaction of X-Ray
radiation with a material is characterized by the refractive index n equal
to unity for vacuum and less than unity for most materials. It can be
written as \cite{PHYSIC}:

\begin{equation}
\label{eq6}
n = 1 - \delta - i\beta
\end{equation}

\noindent
where $\delta $ is the real part of the refraction index deviation from
unity and reflects the material polarization degree upon X-Ray excitation.
The imaginary part $\beta $ characteristics the degree of radiation
attenuation in the material. Values of $\delta $ and $\beta $ for various
media are listed in \cite{TOTAL} and do not exceed $1\cdot 10^{-5}$. The
polarization factor magnitude is directly connect with the critical angle of
total X-Ray beam reflection $\theta _{c}$ \cite{PHYSIC}:

\begin{equation}
\label{eq7}
\delta = \frac{\theta _c^2}{2}
\end{equation}

\noindent
and can be expressed by the wavelength of incident radiation
$\lambda_0$ and common material parameters \cite{PHYSIC}:

\begin{equation}
\label{eq8}
\delta = \frac{{e^2{N}Z'\rho \lambda ^2}}{{2\pi mc^{2}A}}
\end{equation}

\noindent
where $e$ and $m$ are the charge and the mass of an electron, $c$ is the light
velocity, $N$ is the Avogadro number, $Z'$ and $A$ are the effective
charge and atomic mass for the reflectors material, and $\rho $ is its
density. The attenuation factor can be expressed by the linear coefficient
of material absorbtion $\mu $ \cite{PHYSIC}:

\begin{equation}
\label{eq9}
\beta = \frac{\lambda}{4\pi}\mu
\end{equation}

In the conjunction of the secular reflection for an X-Ray radiation the amplitude relation between electric
vectors of the incident and reflected waves and the phase expression have
the from \cite{PHYS2,PHYSIC,TOTAL}:

\begin{equation}
\label{eq10}
\left| {\frac{{\vec {E}_{R}} }{{\vec {E}_{0}} }} \right|_{ \bot}  =
\frac{{\theta - \sqrt {\theta ^{2} - 2\delta - 2i\beta} } }{{\theta + \sqrt
{\theta ^{2} - 2\delta - 2i\beta} } }
\end{equation}

\noindent
and

\begin{equation}
\label{eq11}
\left.
\begin{array}{rccl}
 \cos\psi & = & 2\left( \frac{\theta}{\theta_c} \right)^2 - 1
& \mbox{for}\; \theta \le \theta _{c} \\
 \cos\psi & = & 1 &  \mbox{for}\;\theta > \theta _{c} \\
\end{array}
\right\}
\end{equation}

The X-Ray radiation intensity in the vacuum over the boundary surface in the
interference field area usually defined as
$\left| {\vec A_0 + \vec A_R} \right|^2$
can be presented by the next formula \cite{PHYS2}:

\begin{equation}
\label{eq12}
I\left( {\theta ,z} \right) = \left| {\vec {E}_{0}}  \right|^{2}\left[ {1 +
R + 2\sqrt {R} \cos\left( {\psi - \frac{{2\pi z}}{{D}}} \right)} \right]
\end{equation}

\noindent
where $D$ is the standing wave period along the z-coordinate, expressed by
Eqn.(\ref{eq3}) and R is the reflectivity factor which is a complicated function of
the incident angle $\theta $ \cite{PHYSIC}:

\begin{equation}
\label{eq13}
R = \left| {\frac{{\vec {E}_{R}} }{{\vec {E}_{0}} }} \right|_{ \bot} ^{2} =
\frac{{\left( {\theta - a} \right)^{2} + b^{2}}}{{\left( {\theta + a}
\right)^{2} + b^{2}}}
\end{equation}

\noindent
where $a$ and $b$ are determined by the expressions:

\begin{equation}
\label{eq14}
\left. {
\begin{array}{rcl}
a^{2} & = &
\frac{1}{2}\left[
{\sqrt {(\theta^2 - 2\delta)^2 + 4\beta^2} + (\theta^2 - 2\delta)}
\right] \\
 b^{2} & = &
\frac{1}{2}\left[
{\sqrt {(\theta^2 - 2\delta)^2 + 4\beta^2} - (\theta^2 - 2\delta)}
\right] \\
\end{array}
} \right\}
\end{equation}

The interference field can be observed both at $\theta <\theta _{c}$ and
$\theta >\theta _{c}$, but in the latter angle range its intensity
decreases abruptly \cite{TOTAL}. The standing wave period D achieves its minimum at
$\theta =\theta _{c}$ in the $0 \le \theta  \le \theta _{c}$
range. As the incident angle decreases, the period D increases to become
infinitely large at the grazing incident angle ($\theta =0$). But this is
not case in practice, because the coherence of incidence and reflected beams
is broken. The most obvious factor causing the interference field erosion is
the width finiteness of incident radiation lines $\Delta \lambda$ \cite{PRINCIPLES}.
It is generally accepted, that the interference field does not become smeared,
if the condition \cite{PHYSIC}:

\begin{equation}
\label{eq15}
\Delta \lambda \le \frac{\lambda}{4}
\end{equation}

\noindent
holds.

Another factor influencing the interference field picture is the roughness
of a reflecting surface. Evaluations show that the interference field does
not undergo smearing at $\theta =\theta _{c}$ when the height of
microheterogeneities on the reflected surface does not exceed the critical
size $h_c$ \cite{PHYSIC}:

\begin{equation}
\label{eq16}
h_{c} = \frac{{\lambda} }{{8\sqrt {2\delta} } } = \frac{{1}}{{8}}\sqrt
{\frac{{\pi mc^{2}A}}{{e^{2}N\rho {Z}'}}}
\end{equation}

The critical roughness parameter does not depend on the wavelength of the
incident radiation. Its magnitude for polished optical quartz plates is
equal to 5~nm.

\section{Standing wave formation in the slit of an X-Ray planar waveguide}

A successive reflection of a plane electromagnetic wave in the slit formed
by two parallel plates results in the formation of several interference
field areas in it (Fig.~1b). Varying the slit size can lead to overlapping
the areas to create the uniform interference field zone within the total
clearance of the slit (Fig.~1c)\footnote{It is very important to notice that the uniform interference field zone will be appear, in the context of the specular reflection model, for the some specific reflection angles, only. The uniform interference field zone appearance for any reflection angle can be obtained by taking into consideration the expansion of the interference field into top layers of reflectors and the Goos-Hanchen displacement.}. So, it can be expected that an XSW
excitation can be formed in the plane slit, when it width falls within a
certain range. The minimum slit size for this can be evaluated from
expression (\ref{eq3}) for the critical angle of total reflection:

\begin{equation}
\label{eq17}
D_{min} = \frac{{\lambda} }{{2\sin\theta _{c}} } \approx \frac{{\lambda
}}{{2\theta _{c}} } = \sqrt {\frac{{\pi mc^{2}A}}{{2e^{2}N{Z}'\rho} }}
\end{equation}

The minimum slit size promoting the XSW formation is independent
of an X-Ray incident radiation wavelength. The material structure
density $\rho $ of reflectors is a real factor influencing the
minimum size $D_{min}$. Its magnitude for quartz reflectors
$\left( {\bar Z = 10;\;\bar A = 20} \right)$ is 21~nm. In practice
the $D^*_{min}$ value is smaller because an XSW is characterized
by a visible intensity up to the X-Ray penetration depth $z_e$
into a top layer of a target material, defined by expression
\cite{PHYSIC}:

\begin{equation}
\label{eq18}
\left( {z_{e}}  \right)^{2} = \frac{{\lambda ^{2}}}{{8\pi ^{2}}} \cdot
\frac{{1}}{{\sqrt {\left( {\theta _{c}^{2} - \theta}  \right)^{2} + 4\beta
^{2}} + \left( {\theta _{c}^{2} - \theta ^{2}} \right)}}
\end{equation}

The average value of the penetration depth parameter for an X-Ray slitless
collimator introduced in the work \cite{IZV1E} as
$\bar z_e = z_e  \left( \frac{\theta _c}{2} \right)$
 is 3.6~nm for MoK$_\alpha $ radiation impinging on a quartz plate.
So, the real size of the minimum clearance between the quartz
plates $D^{*}_{min}$ is found to be approximately 14~nm. This value is comparable sized wuth the double roughness of reflectors 10~nm.

The upper restriction for the slit in an X-Ray waveguide can be evaluated
using ratio Eqn.(\ref{eq15}):

\begin{equation}
\label{eq19}
D_{max} = \frac{{\lambda} }{{4\left( {\frac{{\Delta \lambda} }{{\lambda} }}
\right)}} = \frac{{\lambda ^{2}}}{{4\Delta \lambda} }
\end{equation}

For MoK$_\alpha $ radiation, the wavelength is $\lambda =0.707
\cdot 10^{-1}$~nm and $\Delta \lambda =0.29 \cdot 10^{-4}$~nm
\cite{MIRKIN1E}. Substituting these values into Eqn.(\ref{eq19}) gives
$D_{max}=43$\footnote{The important practical parameter
influencing on the upper slit size is a degree of the plate
parallelity in an X-Ray waveguide.}~nm. In practice, its magnitude
is $D^*_{max}=36$~nm. Similar value for CuK$_\alpha $ radiation is
equal to $D^*_{max}=91$~nm. Analogous parameters calculated for
the set of X-Ray and gamma radiation are collected in Table~1.
Quartz total reflection parameters peculiar to the radiation set
are represented in the same place. It should be stressed that the
X-Ray spectra in incident and emergent beams do not agree between
them. For example, if we shall try to form the X-Ray beam excited
by X-Ray tube with Mo anode by using of an X-Ray waveguide with a
slit size $D>D_{max}$, we shall not find the characteristic deposit
in the spectrum of an emergent beam. But if the size will be not excel
the value $D_{max}$ ($D_{min}<D<D_{max}$), the characteristic
radiation MoK$_\alpha$ will be present in the emergent beam X-Ray
spectrum with a great intensity. It can be expected that 
the slit width varying in a waveguide would cause the X-Ray spectrum
modification in the emergent beam at the transportation of the
white X-Ray radiation.

Minimum slit sizes presented in Table~1 (without correction on the
X-Ray depth penetration) are the constant. Maximum values of this
parameter are near 100~nm for the K$_\alpha$ radiation of Fe group
elements. $D_{max}$ and $D_{min}$ magnitudes approach each other
for the hard X-Ray radiation (MoK$_\alpha$, AgK$_\alpha$). The
greatest values of $D_{max}$ for the gamma radiation have engaged
our attention primarily. A significant difference between
$D_{max}$ values for X-Ray and nuclear radiations presents a
unique possibility for its separation in complicate spectra, even
though the radiation wavelengths coinside.

The absolute values of slit widths obtained by calculations have attracted a some attantion. The average value of a slit width accommodating an XSW is
only by an order of magnitude higher than the total surface roughness in an
X-Ray waveguide. If taking also into account that polished surfaces are
characterized by macroheterogeneities, the a high efficiency of a slitless
collimator application for TXRF analysis become evident
\cite{ELPROM1E,IZV1E,ADV1,POV1E}.

\section{Attenuation of an X-Ray standing wave in a planar slit waveguide}

The excitation of a standing wave in a waveguide slit gives rise
to a stationary distribution of the interference field intensity
both along the slit channel (along axis $x$) and crosswise (along
axis $z$). This distribution is pictured in Fig.~2 to fit the
input of a waveguide ($x=0$). The distribution is plotted for a
plane X-Ray beam (CuK$_\alpha $) impinging into a slit of the
quartz waveguide under an angle $\theta=0.92\cdot\theta_c$ on the
reflector surface. The reflection conditions correspond to a value
phase variation $\psi \simeq 45^{\circ}$ and a standing wavelength
$D \approx 1.1 \cdot D_{min}$.  The arising standing wave is
characterized by the penetration depth $z_e=8.6$~nm. Hence, the
distribution shown in Fig.~2 obeys the expression $z_e \approx 0.4
\cdot D_{min}$.

The standing wave intensity within the slit is described by expression (\ref{eq12}).
Outside the slit, the intensity decreases with decrement $\frac{1}{z_{e}}$:

\begin{equation}
\label{eq20}
J\left( {z} \right) = I\left( {\theta _{c} ;z} \right)e^{ - {{z}
\mathord{\left/ {\vphantom {{z} {z_{e}} }} \right.
\kern-\nulldelimiterspace} {z_{e}} }}
\end{equation}

\noindent where I($\theta _{c};z$) is the undisturbed function of
a standing wave intensity defined by expression (\ref{eq12}).
Integration of Eqn.(\ref{eq12}) and Eqn.(\ref{eq20}) produces the
total intensity for a standing wave in the cross-section of an
X-Ray waveguide. The domain of integration for expression
(\ref{eq12}) is equal to a slit size. The standing wave total
intensity in the reflector top layers can be calculated, to a
first approximation, by integrating of function Eqn.(\ref{eq20})
in the top layer domain $1.5\cdot D_{min}$ for each reflector.
Calculations for the CuK$_\alpha $ radiation in the quartz
waveguide show that the total intensity concentrated in the slit
is equal to $L(0)\sim 9.2\;E_0^2D_{min}$ and the total intensity
connected with top layers of the reflectors is $M(0) \sim 1.2
\cdot E_0^2D_{min}$. The standing wave propagation along the slit
channel of a waveguide is characterized by retaining the energy
relation between a slit and top layers of the reflectors. It means
that the relation between the energy in top layers and the total
energy of a standing wave holds too:

\begin{equation}
\label{eq21}
\alpha \left( x \right) = \frac{{M\left( {x} \right)}}{{L\left( {x}
\right) + M\left( x \right)}} = const
\end{equation}

Because attenuation of a standing wave occurs only by the absorbtion in
reflector top layers, equality (\ref{eq21}) implies continuous energy transfer
between different standing wave parts. The attenuation of a standing wave
intensity can then be described by the expression:

\begin{equation}
\label{eq22}
W\left( {x} \right) = \left[ {L\left( {0} \right) + M\left( {0} \right)}
\right]e^{ - \alpha \mu x}
\end{equation}

The magnitude of $\alpha $ depends on the wavelength of incident
radiation, the reflector material properties, the angle of
radiation impinging and the width of a waveguide slit. The
dependence of $\alpha $ on the incident angle can be evaluated by
calculating its variation with some incident angles area of
CuK$_\alpha $ radiation in the quartz waveguide. This gives the
values: $\alpha (\theta _c)=0.8; \alpha(\theta_c/2)=0.05$. The
value of $\alpha $ decreases abruptly, if the slit width exceeds
the wavelength of a standing wave.

Using formula (\ref{eq22}), the values of $\alpha $ for the
incident angle $\theta =0.92\theta_c$ and the condition $s_{slit} \approx 4D$, we one can calculate the total intensity attenuation
for a standing wave of CuK$_\alpha $ radiation in the planar
quartz extensible waveguide. The total intensity of a standing
wave after passing the way $\Delta x=10$~mm in the waveguide is
$W(\theta _{c})=0.3\;W_{0}$ and $W(\theta _{c}/2)=0.6\;W_{0}$.  Note that the model of
multiple total successive reflection under similar conditions
gives: $W_0 \cdot 10^{-27}$ and $0.006\;W_{0}$,
respectively.

The calculated data for the total intensity attenuation of an XSW in a
waveguide lend an explanation of the high efficiency of X-Ray slitless
collimators for X-Ray radiation transport over long distances
\cite{IET2E,IZV1E,ADV1,POV1E}.
However, the evaluations of clearance sizes between the mated quartz plates
in those works strongly disagree ($D \simeq 30$~nm
\cite{ADV1}, $D\sim 150$~nm
\cite{IET2E}). Moreover, the slit width in the collimators is not a stable
parameter and can vary along the collimator length. Therefore, the
evaluations of quantity parameters of the emergent beam intensities for a slitless collimator must be treated with caution. In addition, note that the slit width variation can bring about a considerable
modification of the X-Ray spectrum of an emergent beam, if the initial X-Ray radiation is a mixed type (white and characteristics).

The above results help to elucidate the advantages and shortcomings of X-Ray
slitless collimators. Moreover, they can become the basis for an
X-Ray waveguide designing with properties predicted and high intensity of an emergent
beam.

\section{Application aspects of a planar X-Ray radiation waveguide}

An X-Ray slitless collimator is an X-Ray planar waveguide with an
uncontrolled size of the waveguide slit, which, can be varied during one
experiment. Although these variations are not great, an X-Ray slitless
collimator should be regarded as a simple and convenient experimental model.
For practical purposes, slit waveguides are needed with the greatest possible slit size
for a chosen wavelength. Quartz waveguides with slit sizes 36~nm
and 91~nm are best suited for MoK$_\alpha $ and CuK$_\alpha $ radiations,
respectively. To manufacture such waveguides, the metal thin strips are
deposited on to edges of one quartz reflector of a waveguide, and then the
waveguide is uniformly compressed between metal plates. The clearance
magnitude of a slit can be controlled by the Optical Attenuated Total
Reflection method \cite{INTERNAL}.

An X-Ray slitless collimator came into use for X-Ray fluorescence analysis
of plane surfaces and thin films (TXRF-SC analysis) some years ago
\cite{IZV1E}.
The substitution of the slitless collimator by an X-Ray waveguide
considerably increases X-Ray radiation density on the surface of a target
analyzed and ensures the diagnostic reproduction. Furthermore, an X-Ray
waveguide helps avoid the target contact with reflector surfaces. Otherwise,
an X-Ray planar waveguide retains all advantages of a slitless collimator in
TXRF spectroscopy.
But it is notice that the spurious peaks do not disappear of the waveguide application.

Another important field of a planar waveguide application is in X-Ray
diffraction research. It can be used for structure investigation of
monocrystal surfaces and epitaxial films at the total reflection of an
incident X-Ray beam in the parallel and perpendicular geometries \cite{AP1}
because a waveguide ensures high X-Ray radiation density in the beam. One
more application is for X-Ray energy-dispersive diffractometry of
polycrystal thin films because this waveguide is a beautiful former of a
narrow high intensity beam of the ``white'' X-Ray radiation. The planar
X-Ray waveguide holds greatest promise for using in commercial
diffractometers for symmetrical and asymmetrical geometries.

Special choice of materials for the reflectors top layers is liable to
provide pseudo-monochromatization of X-Ray radiation in an emergent beam.
The waveguide-X-Ray concentrator may be preparated by a surface implantation of reflectors plates with variation of high number ion concentration (Pb, Bi) alone of its length ($n$ -- variation in top layers). 

\section{Planar X-Ray Waveguide -- Monochromator}

Preceding sections are devoted to discussion of the planar X-Ray
waveguide using the total X-Ray reflection phenomenon for the
standing wave generation in a slit space of the device.
But it is well known that the excitation of X-Ray standing wave is possible
for the Bragg geometry too \cite{Phys3}. The X-Ray standing wave arises in
a slit space between two parallel polished reflectors, if the reflectors
are perfect monocrystal been orientated mutually.
The mechanism of the standing wave formation in Bragg reflection conditions
is not practically differed from one in the case of a total X-Ray reflection
(Fig.~1).
But a formal consideration of a standing wave formation in Bragg planar
X-Ray waveguide (waveguide-monochromator) requests
replacement of the X-Ray depth penetration parameter $z_e$ on the parameter
of the primary extinction length $z_{ext}$ \cite{SSP1}:

\begin{equation}
\label{eq23}
z_{ext}
=
\frac{1}{\sigma}
=
\frac{\sin\theta_b}{2\lambda|c|}
\cdot
\frac{m c^2}{e^2}
\cdot
\frac{v}{|F_h|}
\end{equation}

\noindent
where $\sigma$ is the extinction factor, $c$ is the polarization factor
been equal to unity for a $\sigma$-polarization,
$c$ is the light velocity,
$m$ and $e$ are the electron characteristics, $v$ is the unit cell volume for the reflector material, $F_h$ is the relative structure factor of the
chosen reflection.
The $z_{ext}$ magnitude defines the crystal thickness attenuated of the
X-Ray beam intensity falling on the crystal under Bragg angle
with <<$e$>> factor.
It is significantly that the primary extinction length is not depended
on a wavelength of an X-Ray radiation.
Magnitudes of the primary extinction length are usually two orders higher
as values of the depth penetration at the X-Ray total reflection.
For example, the magnitude of $z_{ext}$ for (200) NaCl reflection is
equal to 660~nm \cite{SSP1}.
Because of this, the practical size of a waveguide-monochromator slit may
be visibly differed from one calculated and its efficiency must be lower
as compared to a waveguide built on the total reflection phenomenon.
But the X-Ray radiation density in its emergent beam will be significantly
higher as one for the conventional Bonse-Hart monochromator.
\textbf {\emph{The planar X-Ray waveguide-monochromator application is conceptually identical with the
Borrmann effect manifistation in perfect crystals \cite{PhysZ1, X-Ray2}.}}
Authors hope to give a comprehensive analysis of peculiarities which are
typical for the planar X-Ray waveguide-monochromator in other work.

\section{Conclusion}

The model of XSW excitation in a planar slit waveguide has been developed by
employing the interference wave theory to treat areas whose size
considerably exceeds the wavelength of initial radiation but the coherence
conditions are still valid. An example of practical embodiment of the idea
of such a waveguide is an X-Ray slitless collimator whose unique properties
could only be explained in terms of the model of XSW excitation. The
evaluation of the upper and lower boundaries for the slit width which
provide XSW excitation, points out the way to waveguides building with most
efficient for particular purposes.

The presented model is a simple one and disregards some phenomena as such existence of the X-Ray interference field in top layers of waveguide reflectors, 
the Goos-Hanchen effect, monotonous modification of the refraction index on
a vacuum-material interface \cite{PhSol1} and others. However, even the simplified
evaluations made in the work show that a planar X-Ray waveguide and a waveguide-monochromator can became useful tools for the X-Ray diffraction and spectroscopic investigations, especially for the work with synchrotron radiation.

\textbf {\emph{It is very important that the model can be applicated for the neutron and electron beams and will stimulate the waveguides creation both for the white radiation and for the monochromatic one for them.}} The waveguide-monochromator can be basis for the building of a laser pumping system applicated for the hard X-Ray and gamma radiation. The idea of the simplest X-Ray accumulator functioned on base of the waveguide-monochromator will be described in another place.

Direct experiments with planar X-Ray waveguides is carried out today and will be published soon after.

\section{Acknowledgements}

The authors thank d-r A.V.~Okhulkov for useful discussion of the problems and O.S.~Kondratiev for the help in calculations.

\newpage

\renewcommand{\refname}

\newpage
\section{Captions to the Figures}

\hangindent=2cm \noindent Figure~1.~Classical scheme of the
standing wave field formation at a specular reflection of a plane
monochromatic wave above the mirror surface (a), a scheme of the
standing wave field formation at multiple successive reflection of
a plane monochromatic wave (b), the principle scheme explaining
the formation of the standing wave uniform zone upon trapping
monochromatic plane wave radiation by a planar extensive waveguide
(c).
\bigskip

\hangindent=2cm \noindent
Figure~2.~Intensity distribution
function for an XSW in a waveguide slit and a top layer of quartz
reflectors for an X-Ray beam impinging on the slit under a
certain angle of total reflection $\theta$ for the quartz reflectors. The
function without absorbtion upon total reflection is shown by a
dashed line. The function reflects the picture for
$\lambda=0.1541$~nm (CuK$_\alpha$); $\theta=0.92 \cdot \theta_c$;
$s=97$~nm.

\end{document}